\documentclass[aps,prb,twocolumn,floatfix,amsmath,amssymb,superscriptaddress]{revtex4}

\usepackage{graphicx} 

\begin{document}
%---------------------------------------------------------------------
%

\title{Steam reforming on transition-metal carbides from
  density-functional theory}
 
\author{Aleksandra Vojvodic}
\affiliation{Department of Applied Physics, Chalmers University of
  Technology SE-412 96 G\"{o}teborg, Sweden}   
\email{alevoj@chalmers.se}

%
%%%%%%%%%%%%%%%%%%%%%%%%%%%%%% ABSTRACT %%%%%%%%%%%%%%%%%%%%%%%%%%%%%%
%

\begin{abstract} 
A screening study of the steam reforming reaction 
($\text{CH}_4 + \text{H}_2\text{O}\rightarrow \text{CO} + 3\text{H}_2$) 
on early transition-metal carbides (TMC's) is performed by means of
density-functional theory calculations. The set of considered surfaces
includes the $\alpha$-Mo$_2$C$(100)$ surfaces, the low-index ($111$)
and ($100$) surfaces of TiC, VC, and $\delta$-MoC, and the oxygenated
$\alpha$-Mo$_2$C$(100)$ and TMC($111$) surfaces. It is found that 
carbides provide a wide spectrum of reactivities towards the steam
reforming reaction, from too reactive via suitable to too inert. The
reactivity is discussed in terms of the electronic structure of the
clean surfaces. Two surfaces, the $\delta$-MoC$(100)$ and the oxygen
passivated $\alpha$-Mo$_2$C$(100)$ surfaces, are identified as
promising steam reforming catalysts. These findings suggest that
carbides provide a playground for reactivity tuning, comparable to the
one for pure metals.    
\end{abstract}

%
%%%%%%%%%%%%%%%%%%%%%%%%%%%%%% ABSTRACT %%%%%%%%%%%%%%%%%%%%%%%%%%%%%%
%

\maketitle

%
%%%%%%%%%%%%%%%%%%%%%%%%%%% INTRODUCTION %%%%%%%%%%%%%%%%%%%%%%%%%%%%%
%

\section{Introduction}

Steam reforming is an important industrial process, where natural gas
(CH$_4$) is converted into synthesis gas (CO and H$_2$) according to
the overall reaction  
\begin{equation}
\text{CH}_4 + \text{H}_2\text{O} \rightleftharpoons \text{CO} + 3\text{H}_2. 
\end{equation}
The synthesis gas is subsequently transformed into more valuable
chemicals, such as ammonia, methanol, and diesel. Since steam reforming
acts as a source of hydrogen, it is also potentially important for any
emerging hydrogen economy. For a detailed review of the steam
reforming process and its applications the reader can consult
Ref.~\onlinecite{RostrupNielsen02}.       

Commercially the steam reforming reaction is conducted over a Ni-based
catalyst due to the relatively low cost and good activity of
nickel. This reaction has been studied in detail on the close-packed 
Ni($111$) surface and on the stepped Ni$(211$) surface by means of
density-functional theory.\cite{Bengaard02} The major technological
challenge for Ni catalysts is the formation of carbonaceous deposits,
termed coke, that lead to catalyst deactivation. In
Refs.~\onlinecite{Bengaard02,Helveg04} it was established that the
step edges on Ni surfaces act as growth centers for graphite. Other
transition metals (TM), such as Ru, Rh, Pd, Ir, and Pt also show high
activity and selectivity towards steam reforming\cite{Jones08} and
have a high resistance against carbon formation. However, these
materials are scarce in nature and expensive. Therefore new materials
that are resistant to carbon formation are needed.            

Transition-metal carbides (TMC's) have gained quite some attention
since Levy and Boudart reported that they have "platinum-like
behavior'' for certain reactions.\cite{LeBo73} The starting material
for the production of carbides is cheap and abundant and therefore it
has been suggested that they can replace the noble metals in
catalysis. The main problem with the carbides has been to produce 
materials with sufficiently high surface area for them to be
interesting for catalytic applications. This problem has been overcome
and several studies report on carbides with surface areas as high as
200~m$^2/$g.\cite{Volpe85,Lee87,Lee92,Oyama96}         

In Refs.~\onlinecite{York97} and \onlinecite{Claridge98}, it was shown
that carbides of molybdenum and tungsten are stable and extremely
active catalysts for not only the steam reforming but also the dry
reforming and the partial oxidation of methane at elevated
pressure. Moreover, no macroscopic carbon was deposition on the
catalysts during the catalytic reactions. The relative activity of a
number of steam reforming  catalysts was established as:       
Ru $>$ Rh $>$ Ir $\approx$ Mo$_2$C $>$ WC$>$ Pd $>$ Pt.  

Several density-functional theory studies concerning adsorption on
TMC's have been conducted, for example,  
atomic adsorption,\cite{Zhang,Kitchin05,VinesTMC100,VoRuLu06,RuVoLu06,RuLu07,RuVoLu07,VoRuLu09,VoHeRuLu09}   
O$_2$ adsorption,\cite{VinesO2TMC100}
CO adsorption,\cite{Didziulis,LiuCOTiCMo2C, Ren05} 
NH$_x$ ($x=1-3$) adsorption,\cite{VoRuLu09,VoHeRuLu09}  
methanol adsorption,\cite{Pistonesi08} 
CH$_x$ ($x=0-3$) and C$_2$H$_4$ adsorption.\cite{Ren} 
In addition, the water-gas-shift reaction has been studied in
Refs.~\onlinecite{TominagaWGS,LiuWGS,VinesWGS}. 

To our knowledge, a theoretical understanding of the steam reforming 
reaction on TMC's is missing in the literature. The aim of this paper
is to investigate how the TMC's are suited for the steam reforming
reaction by performing a set of density-functional theory (DFT)
calculations on different surfaces of $\alpha$-Mo$_2$C, TiC, VC, and
$\delta$-MoC in order to catch the trends between TMC's composed of
different TM atoms and of different structures. The influence of the 
surfaces being covered with oxygen is also investigated. 

The outline of this paper is as follows. To begin with the systems to
be investigated are defined in Sec.~\ref{sec:systems}. This is
followed by Sec.~\ref{sec:calculations}, which describes the
calculation details. Our results are presented in
Sec.~\ref{sec:results} and discussed in
Section~\ref{sec:discussion}. The paper ends with a conclusion in
Section~\ref{sec:conclusion}.

%
%%%%%%%%%%%%%%%%%%%%%% THE CARBIDE SYSTEMS %%%%%%%%%%%%%%%%%%%%%%%%%%  
%
\section{Investigated TMC systems}\label{sec:systems}

\begin{figure*}
\centering
\includegraphics[width=\textwidth]{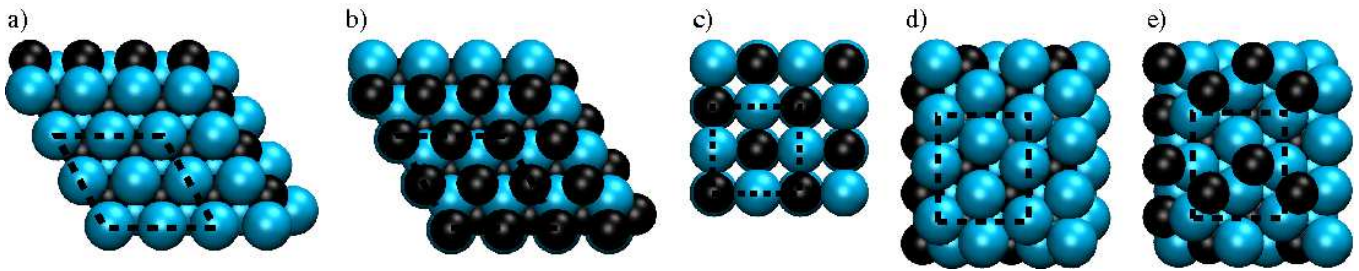}
\caption{\label{fig:structure}
Top view of the relaxed 
(a) Mo-terminated $\delta$-MoC($111$), 
(b) C-terminated $\delta$-MoC($111$), 
(c) $\delta$-MoC($100$), 
(d) Mo-terminated $\alpha$-Mo$_2$C($100$), 
(e) C-terminated $\alpha$-Mo$_2$C($100$) surface. The blue and
black balls represent the Mo and C atoms, respectively. The black
dashed lines indicate the unit cell.} 
\end{figure*}

The Mo--C system can occur in several crystalline forms, see for
example the phase diagrams in Refs.~\onlinecite{Dubois88} and
\onlinecite{VeKuKu88}. There are two types of Mo$_2$C phases, one
orthorhombic ($\alpha$) and one hexagonal ($\beta$), which is a high
temperature phase. In this study, we focus on the $\alpha$ phase. For
this phase the Mo atoms are only slightly distorted from an hcp
arrangement and the carbon atoms occupy one half of the octahedral
interstitial sites. In the present molybdenum carbide literature there
seems to be a confusion when it comes to the attribution of the
symbols to the different phases, especially regarding the $\alpha$ and
$\beta$ phases. This is probably because the $\alpha$-Mo$_2$C phase
was originally assigned a hexagonal structure but refined experiments
identified the structure to be of orthorhombic type\cite{Parthe63},
with lattice parameters $a=4.729$~\AA, $b=6.028$~\AA \ and
$c=5.197$~\AA.\cite{Otani95}     

The close-packed surface of $\alpha$-Mo$_2$C is the ($100$)
surface. It consists of alternating Mo and C layers and can be either
Mo- or C-terminated. Experimental studies of $\alpha$-Mo$_2$C($100$)
show that the surface termination and structure depend strongly on the
cleaning procedure. Both the Mo- and the C-terminated surfaces have
been observed.~\cite{Clair99,Clair00} Therefore we consider both the
TM- and the C-terminated carbide surfaces.           

The TiC, VC and $\delta$-MoC adopt the NaCl structure. The ($100$)
surface of these carbides consists of layers of equal amounts of TM
and C atoms. The ($111$) surface, however, is polar and consists of
alternating TM and C layers, similar to the $\alpha$-Mo$_2$C($100$)
surfaces. One important difference is that the C-terminated
$\alpha$-Mo$_2$C($100$) surface has half the amount of C atoms as
compared to a C-terminated TMC($111$) surface.    

In this paper, we report the results from DFT calculations on 
(i) the TM- and C-terminated surfaces of $\alpha$-Mo$_2$C($100$),  
TiC($111$), VC($111$), and $\delta$-MoC($111$); 
(ii) the O-covered surfaces in (i);
and 
(iii) the ($100$) surfaces of TiC, VC and $\delta$-MoC.

%
%%%%%%%%%%%%%%%% COMPUTATIONAL DETAILS %%%%%%%%%%%%%%%%%%%%%%%%%%%%%
%
\section{Theoretical details}\label{sec:calculations}

All the DFT calculations in this paper are performed
with the computer code DACAPO\cite{dacapo}, which uses plane-waves and 
ultra-soft pseudopotentials. For the optimization of the bulk lattice
parameters of the considered TMC's a plane-wave cutoff of $400$~eV, a
$8\times8\times8$ $k$-point sampling, and the RPBE
exchange-correlation (\textit{xc}) functional\cite{Hammer} are
used. The bulk structure of $\alpha$-Mo$_2$C was obtained by
minimizing the total energy of the unit cell with respect to the
length of the lattice vectors using a Newton-Raphson scheme and
allowing for complete internal relaxation at each step. The calculated
lattice parameters $a=4.825$~\AA, $b=6.162$~\AA, and $c=5.304$~\AA
\ differ by $\sim 2\%$ from the experimental values.\cite{Otani95}         

For TiC, VC and $\delta$-MoC, the equilibrium bulk structure was
obtained utilizing the Murnaghan equation of state.\cite{FuHo83} The
calculated lattice constants are $4.336$~\AA, $4.171$~\AA, and
$4.444$~\AA \ for TiC, VC, and $\delta$-MoC, respectively. A
quantitative agreement is found between the theoretical and
experimental lattice constants.\cite{NaYa05,GuHaGr92}          

The surfaces are modeled using the supercell approach, with the DFT
lattice parameters given above. A $400$~eV energy cutoff for the
plane-wave expansion of the one-electron orbitals, a $4 \times 4\times
1$ $k$-point grid, and the RPBE \textit{xc} functional are used. Each
TMC($111$) and TMC($100$) surface is represented by a slab with the
geometry defined in Fig.~\ref{fig:structure} and a thickness of four
atomic layers. Repeated slabs are separated by at least a $10.8$~\AA \
thick vacuum region. The Mo- and C-terminated $\alpha$-Mo$_2$C($100$)
slabs have a surface geometry as defined in Fig.~\ref{fig:structure}
and consist of four layers Mo and four layers C separated by
$10.6$~\AA \ vacuum. During structural relaxations, the two bottom
atomic layers are constrained to the corresponding bulk geometry.          

Figure~\ref{fig:structure} shows the relaxed structures of the Mo-
and C-terminated $\delta$-MoC($111$) surfaces, the $\delta$-MoC($100$)
surface, and the Mo- and C-terminated $\alpha$-Mo$_2$C($100$)
surfaces. These relaxed surfaces and the corresponding VC and TiC
surfaces are used as starting points for the modeling of the steam
reforming reaction.      

The investigated steam reforming reaction is assumed to have the
following nine elementary steps:
\begin{align}\label{eq:reactionstep1}
\text{CH}_4          +2^*        &\rightleftharpoons \text{CH}_3^* + \text{H}^*  \\
\text{CH}_3^*        +^*         &\rightleftharpoons \text{CH}_2^* + \text{H}^*  \\
\text{CH}_2^*        +^*         &\rightleftharpoons \text{CH}^*   + \text{H}^*  \\
\text{CH}^*          +^*         &\rightleftharpoons \text{C}^*    + \text{H}^*  \\
\text{H}_2\text{O} +2^*        &\rightleftharpoons \text{OH}^*   + \text{H}^*  \\
\text{OH}^*          +^*         &\rightleftharpoons \text{O}^*    + \text{H}^*  \\
\text{C}^*           +\text{O}^* &\rightleftharpoons \text{CO}^*                 \\
\text{CO}^*                      &\rightleftharpoons \text{CO}     +^*           \\
2\text{H}^*                      &\rightleftharpoons \text{H}_2    +2^*,          
\label{eq:reactionstep9}
\end{align}
where $^*$ (2$^*$) denotes one (two) empty surface site and $X^*$
denotes the adsorbate $X$ bonded to this site. Adsorption of all the
intermediates (CH$_3$, CH$_2$, CH, C, OH, O, CO, and H) in
Eqs.~(\ref{eq:reactionstep1})--(\ref{eq:reactionstep9}) are
considered. In all adsorption calculations, the adsorbate coverage is
one-quarter of a monolayer (ML). Adsorption is allowed on one of the
two slab surfaces at the time. Adsorbates together with the surfaces
are allowed to relax in all directions until the forces are less than
$0.05$~eV/\AA. The energies in the calculated potential energy
diagrams are expressed relative to the clean surface (O-covered
surfaces if the surface is oxygenated), a CH$_4$ molecule, and a
H$_2$O molecule in the gas phase.              

The steam reforming reaction on the oxygenated TMC surfaces is modeled
by adsorbing the reaction intermediates on a 1ML O-covered TMC
surface. According to DFT calculations in Ref.~\onlinecite{LiuWGS}, the
O covering of the Mo- and C-terminated Mo$_2$C($100$) surfaces is
energetically favorable up to 1ML. Our calculations give the same
structure for the 1ML O-covered Mo- and C-terminated Mo$_2$C($100$)
surfaces as in Ref.~\onlinecite{LiuWGS}.  

%
%%%%%%%%%%%%%%%%%%%%%% RESULTS %%%%%%%%%%%%%%%%%%%%%%%%%%%%%%%%%%%%%%%
%
\section{Results}\label{sec:results}
This Section focuses on the results obtained from the screening
calculations study. We provide the calculated potential energy diagrams
for the steam reforming reaction on all the considered TMC
surfaces. These are compared with existing results on Ni($111$) and
Ni($211$)~\cite{Bengaard02}.

%
%%%%%%%%%%%%%%%%%%%%%%%%%%% RESULTS %%%%%%%%%%%%%%%%%%%%%%%%%%%%%%%%%%
%

\subsection{TMC($111$) and $\alpha$-Mo$_2$C($100$) surfaces}\label{sec:TMC111}   
\begin{figure}
\centering
\includegraphics[width=\columnwidth]{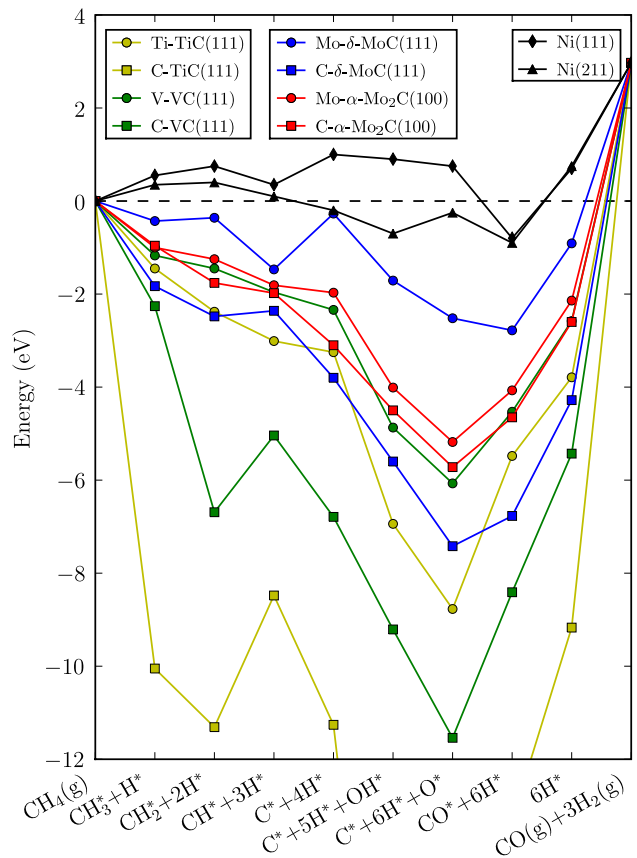}
\caption{\label{fig:TMC111}
Calculated potential energy diagram for the steam reforming reaction
on the TM-terminated (circles) and C-terminated (squares) TMC($111$)
and $\alpha$-Mo$_2$C($100$) surfaces. For comparison the Ni($111$) and
Ni($211$) data, adapted from Ref.~\onlinecite{Bengaard02}, are given.}         
\end{figure}

The calculated potential energy diagram for the steam reforming
reaction on the TM- and C-terminated TMC's is shown in
Fig.~\ref{fig:TMC111}. First of all we notice that the span of the
potential energies of the investigated surfaces is huge. Compared to
the Ni surfaces~\cite{Bengaard02}, all the TMC surfaces are much more
reactive because they bind the intermediates too strongly. Therefore
none of these surfaces is suitable for the steam reforming reaction.  

In the following we present the trends in reactivity that are found
when changing the TM atom in the TMC compound. For each of the
considered TMC's, the C-terminated surface is found to be more
reactive than the TM-terminated one. This is to be expected since
the TMC($111$) surfaces are found to be TM terminated under UHV
conditions.\cite{AonoTiC,RundgrenVC} All the TM-terminated TMC($111$)
surfaces have similar potential energy shape as a function of the
reaction intermediates. The same is found for the C-terminated group
of TMC surfaces. For the TM-terminated TMC surfaces, the reactivity
can be ordered as       
TiC($111$) $>$ VC($111$) $>$ $\alpha$-Mo$_2$C($100$) $>$ $\delta$-MoC($111$),  
while for the C-terminated TMC surfaces we find that the reactivity
order is  
TiC($111$) $>$ VC($111$) $>$ $\delta$-MoC($111$)     $>$ $\alpha$-Mo$_2$C($100$).  
Hence different trends are observed for the different
terminations. 

The considered $\alpha$-Mo$_2$C($100$) surfaces are less reactive than
the C-terminated $\delta$-MoC($111$) surface but more reactive than
the Mo-terminated $\delta$-MoC($111$). For a given carbide, the
smallest difference in reactivity between the TM- and the C-terminated
surfaces is found for $\alpha$-Mo$_2$C($100$). This can be due to the
fact that the C-terminated $\alpha$-Mo$_2$C($100$) surface has $50\%$
less C than the C-terminated TMC($111$) surfaces.     

The largest minimum in the potential energy surface is found upon 
adsorption of O-carrying intermediates, that is, a strong O--surface
bond is formed. This implies that the TMC($111$) surfaces are easily
oxidized. Experimental studies show that the O$_2$ adsorption on 
TiC($111$)\cite{Brandshaw80} and
$\alpha$-Mo$_2$C($100$)\cite{Clair00} are dissociative. Therefore we
investigate how the steam reforming reaction is influenced when the
TMC($111$) and the $\alpha$-Mo$_2$C($100$) surfaces are oxygenated.

\subsection{O-covered TMC($111$) and O-covered $\alpha$-Mo$_2$C($100$)
  surfaces}\label{sec:OTMC111} 

\begin{figure}
\centering
\includegraphics[width=\columnwidth]{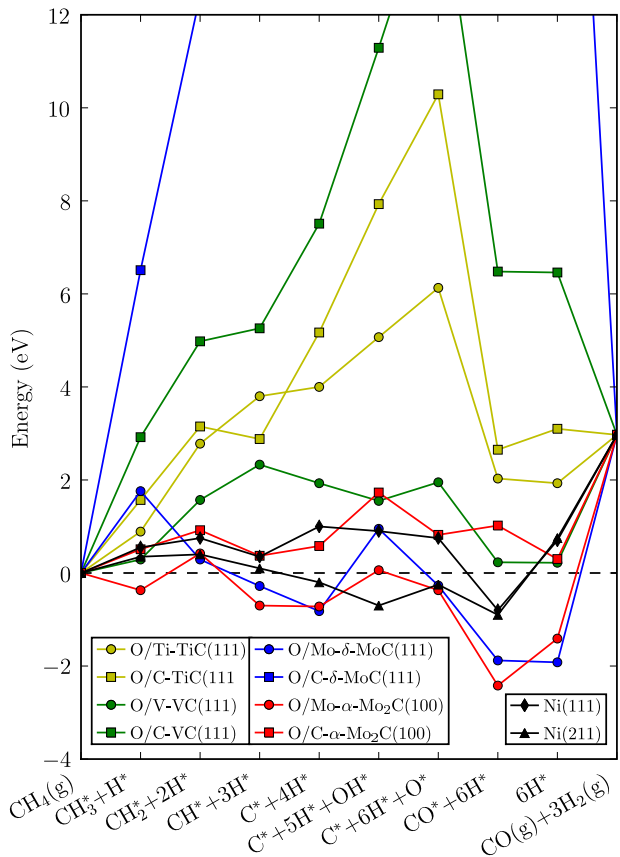}
\caption{\label{fig:OTMC111}
Calculated potential energy diagram for the 1ML O-covered
TM-terminated (circles) and C-terminated (squares) TMC($111$) and
$\alpha$-Mo$_2$C$(100)$ surfaces. Also given are the Ni($111$) and
Ni($211$) data, adapted from Ref.~\onlinecite{Bengaard02}.}            
\end{figure}

Figure~\ref{fig:OTMC111} shows the calculated potential energy diagram
for the steam reforming reaction on the 1ML O-covered TM- and
C-terminated TMC surfaces. We find that the O-covered C-terminated 
$\alpha$-Mo$_2$C($100$) has a potential energy profile similar to that
reported for the Ni($111$) surface, which is the preferred Ni facet 
for the steam reforming reaction according to the DFT calculations in
Ref.~\onlinecite{Bengaard02}. The other O-covered surfaces are
unsuitable as catalysts for the steam reforming either because several
of the reaction steps are significantly uphill energetically, or
because the surface gets poisoned by some of the intermediates.  

Some of the trends that are found for this class of surfaces are here
described. Each of the O-covered TMC surfaces is less reactive than
its corresponding non-oxidized surface. For a given O-covered TMC
surface the TM-terminated surface has higher reactivity than the
C-terminated one, in contrast to the finding on the non-oxidized
surfaces. For the O-covered TM-terminated TMC surfaces, the reactivity
can be ordered as follows          
$\alpha$-Mo$_2$C($100$) $>$ $\delta$-MoC($111$) $>$ VC($111$) $>$ TiC($111$),  
while for the O-covered C-terminated TMC surfaces we find that  
$\alpha$-Mo$_2$C($100$) $>$ TiC($111$) $>$ VC($111$) $>$ $\delta$-MoC($111$).
The order in reactivity between the different surfaces is thus
different for the TM- and for the C-terminated O-covered surfaces.

\subsection{TMC($100$) surfaces}
\begin{figure}
\centering
\includegraphics[width=\columnwidth]{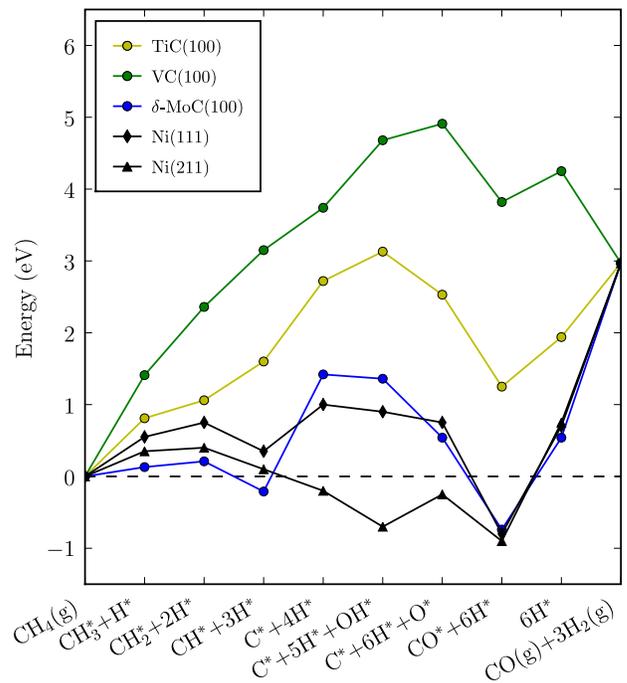}
\caption{\label{fig:TMC100}
Calculated potential energy diagram for TMC($100$). The Ni($111$) and
Ni($211$) data given here are adapted from
Ref.~\onlinecite{Bengaard02}.}      
\end{figure}

The calculated potential energy diagram for the steam reforming
reaction on the TMC($100$) surfaces is shown in
Fig.~\ref{fig:TMC100}. The $\delta$-MoC($100$) surface is found to
have a potential energy curve similar to the one of the Ni($111$)
surface. We find the same shape of the potential energy curves for 
all the TMC($100$) surfaces. A local minimum is found for all the
TMC($100$) surfaces upon adsorption of CO. The reactivity of the
surfaces follows the order 
$\delta$-MoC($100$) $>$ TiC($100$) $>$ VC($100$), that is, a non
linear variation when moving from right to left in the periodic
table. Compared to the TMC($111$) surface the TMC($100$) surfaces 
are far less reactive. In Ref.~\onlinecite{VinesWGS} it was suggested
that the TiC($100$) surface is a suitable catalyst for the
water-gas-shift reaction. However, according to our calculation it is
not the surface of choice for the steam reforming reaction since it is 
inert (see Fig.~\ref{fig:TMC100}).

%
%%%%%%%%%%%%%%%% DISCUSSION %%%%%%%%%%%%%%%%%%%%%%%%%%%%%
%

\section{Discussion}\label{sec:discussion}
The conducted computational screening study clearly shows that only
two surfaces, the O-covered C-terminated $\alpha$-Mo$_2$C($100$) and
the $\delta$-MoC($100$), are suitable for the steam reforming 
reaction. These two surfaces are found to have potential energy
profiles similar to the one of Ni($111$), which is the preferred Ni
surface for steam reforming according to Ref.~\onlinecite{Bengaard02}.  

The obtained potential energy diagrams show that there is a large
variety in the reactivities of the different carbides (see
Figs.~\ref{fig:TMC111},~\ref{fig:OTMC111} and \ref{fig:TMC100}).
However, besides the two candidate surfaces none of the other
considered TMC surfaces are suitable for the steam reforming reaction
since they either get self-poisoned by the intermediates or they show
a significant energetical uphill for some of the elementary steps in 
Eqs.~(\ref{eq:reactionstep1})--(\ref{eq:reactionstep9}).       

An understanding of the reactivity can be achieved by studying the
electronic structure of the surface. In
Refs.~\onlinecite{VoRuLu06,RuVoLu06,RuLu07,RuVoLu07,VoRuLu09,VoHeRuLu09}
a thorough analysis of the electronic structure established that the
TM-terminated TMC($111$) surfaces possess surface resonances
(SR's). These were identified to be responsible for strong
adsorbate-surface bonds. Therefore the high reactivities of the
TM-terminated TMC($111$) surfaces (see Fig.~\ref{fig:TMC111}) are
attributed to the presence of these SR's. Preliminary electronic
structure calculations show that similar resonances are present also
on the layered polar $\alpha$-Mo$_2$C($100$) surfaces, which would
explain their high reactivities.      

The NaCl TMC($100$) surfaces, on the other hand, do not posses any
SR's\cite{Oshima81,RuLu07} and show a much lower reactivity (see
Fig.~\ref{fig:TMC100}) compared to the TMC($111$) surfaces. Since a
suitable candidate, the $\delta$-MoC($100$) surface, is found among
these surfaces we have tried to change the reactivity of the
TMC($111$) surfaces by weakening their SR's.  

The lower reactivities of the O-covered TMC($111$) and
$\alpha$-Mo$_2$C($100$) surfaces (see Fig.~\ref{fig:OTMC111}),
compared to the corresponding clean surfaces can be understood as
follows. Adsorption of oxygen on the TMC($111$) results in a quenching
of the SR's. Thus, the surfaces are passivated, that is, their ability
to form strong bonds with other adsorbates decreases due to the strong
O--surface bond. This result shows that the reactivity of carbides
surfaces can be tuned.      

An important question for catalytic applications is the stability of
the candidate materials. Steam reforming operates over a large range
of working conditions with temperatures $700-1100$~$^{\circ}$C and
pressures $1-25$~bar. The most recent phase diagram for
Mo--C systems is the one by Velikanova \textit{et al.}\cite{VeKuKu88} It
shows that the bulk $\alpha$-Mo$_2$C phase exists for temperatures $<
1440$~$^{\circ}$C, while the $\delta$-MoC is a high temperature phase
which exists above $1956$~$^{\circ}$C. Based only on bulk data,
$\alpha$-Mo$_2$C is preferred over $\delta$-MoC.  

When it comes to the stability of a surface the situation can of
course differ from the bulk. According to York and
coworkers\cite{York97,Claridge98}, Mo$_2$C was active and selective
for the stoichiometric steam reforming of methane to synthesis
gas. The catalyst deactivated at atmospheric pressure, but was very
stable when elevated pressures were employed, and no carbon deposition
was observed on the catalysts. In those studies the surface
characteristics was, however, not addressed. 

A surface characterization of $\alpha$-Mo$_2$C($100)$ has been
performed in Ref.~\onlinecite{Clair99}. The surface preparation
consisted of Ar ion bombardment followed by annealing at successively
higher temperatures. The surface was found to change from being Mo
terminated to C terminated at annealing temperatures above
$1300$~K. Despite that these surfaces are stable at temperatures 
relevant for the steam reforming, our study shows that they are too
reactive. Adsorption of O$_2$ on both the Mo- and the C-terminated
$\alpha$-Mo$_2$C($100)$ surfaces were found to be dissociative and the
surfaces were covered with oxygen at $800$~K.\cite{Clair00} This
supports the stability of the O-covered $\alpha$-Mo$_2$C($100$)
surface. 

In addition, stable crystalline molybdenum carbide films prepared by
chemical vapor deposition (CVD) and physical vapor deposition (PVD)
have been found to contain either Mo$_2$C and/or the $\delta
$-MoC$_{1-x}$ phase.\cite{Wood,Haase,Okuyama,Lu} Hence it seems
plausible that both the O-covered C-terminated $\alpha$-Mo$_2$C($100$)
and the $\delta$-MoC($100$) surfaces may be stable at relevant steam
reforming conditions, however, further studies are needed.

%
%%%%%%%%%%%%%%%%%%%%%% CONCLUSION %%%%%%%%%%%%%%%%%%%%%%%%%%%%%%%%%%%%%%%%%%%%%%
%

\section{Conclusion}\label{sec:conclusion}
In this paper, a computational screening study of the steam reforming
reaction on transition-mental carbides has been presented. By using a
screening approach we are able to map out the huge reactivity space,
which ranges from very reactive to inert, provided by the carbides. A
small set of surfaces, consisting of the $\delta$-MoC($100$) and the
O-covered C-terminated $\alpha$-Mo$_2$C($100$), is identified as
promising candidate for the steam reforming reaction. With this
thermochemical data at hand we lay a basis for a thermodynamic and
micro-kinetic treatment of the carbides. This study illustrates the
versatility and tunabilty of the TMC systems for the steam reforming
reaction. The results can be understood in terms of the previously
proposed model of chemisorption on TMC surfaces, in which surface
resonances play crucial roles. However, further studies of the
surface electronic structure are needed to establish what surface
characteristics that are required of a carbide to make is a suitable
catalyst for steam reforming.

%
%%%%%%%%%%%%%%%%%%%%%% ACKNOWLEDGEMENTS %%%%%%%%%%%%%%%%%%%%%%%%%%%%%%%%%%%%%%%%
%
\section*{Acknowledgments}
The calculations were performed at HPC2N via the Swedish National
Infrastructure for Computing. The author thanks Anders Hellman and
Carlo Ruberto for constructive discussions and comments on the
manuscripts.

%%%%%%%%%%%%%%%%%%%%%%%%%%%% REFERENCES %%%%%%%%%%%%%%%%%%%%%%%%%%%%%
%---------------------------------------------------------------------
%

%
%---------------------------------------------------------------------
%

\end{document}